\documentclass[10pt, a4paper]{article}

\usepackage{graphicx}
\usepackage{amssymb}
\usepackage{amsmath}
\usepackage{amsthm}
\usepackage{latexsym}

\newtheorem{thm}{Theorem}[section]
\newtheorem{proposition}{Proposition}[section]
\newtheorem{deff}{Definition}[section]

\newtheorem{lem}{Lemma}[section]

\newtheorem{problem}{Problem}[section]

\begin{document}
\title{ An explicit family of unitaries with exponentially minimal length Pauli geodesics }
\author{Wei Huang \\ Electrical Engineering and Computer Science Department, \\
University of Michigan, Ann Arbor, MI \\weihuang@eecs.umich.edu }
\maketitle

\begin{abstract}
Recently, Nielsen et al ~\cite{nielsen-2005,nielsen-2006-311,
dowling-2006,nielsen-2006} have proposed a geometric approach to
quantum computation. They've shown that the size of the minimum
quantum circuits implementing a unitary U, up to polynomial factors,
equals to the length of minimal geodesic from identity I through U.
They've investigated a large class of solutions to the geodesic
equation, called Pauli geodesics. They've raised a natural question
whether we can explicitly construct a family of unitaries U that
have exponentially long minimal length Pauli geodesics? We give a
positive answer to this question.
\end{abstract}

\section{Preliminary}

\subsection{Pauli basis and Pauli metrics}

We define \emph{general Pauli matrix} $\sigma$ as tensor product of
identity matrix or Pauli matrices X, Y or Z. We define \emph{pauli
weight} of a general Pauli matrix $\sigma$ as the total number of X,
Y and Z in $\sigma$, noted by pw($\sigma$). Given a control
Hamiltonian H, we can write H in terms of the Pauli operator
expansion $H = \sum\limits_\sigma{\lambda_{\sigma}}\sigma +
\sum\limits_\tau{\lambda_{\tau}}\tau$, where in the first sum
$\sigma$ ranges over all possible one and two-body interactions with
$pw(\sigma)\leq 2$ , while the second sum $\tau$ ranges over all
general Pauli matrix $\tau $ with $pw(\tau)\geq 3$. We then define
several \emph{Pauli metrics} as following:

\[ F_q(H) =
\sqrt{\sum\limits_\sigma{\lambda_{\sigma}^2} + q^2
\sum\limits_\tau{\lambda_{\tau}^2}} \].

\[ F_2(H) =
\sqrt{\sum\limits_\sigma{\lambda_{\sigma}^2} +
\sum\limits_\tau{\lambda_{\tau}^2}} \].

\[ F_1(H) =
\sqrt{\sum\limits_\sigma{|\lambda_{\sigma}|} +
\sum\limits_\tau{|\lambda_{\tau}|}} \].

\subsection{Geodesic Equation}

A \emph{geodesic} is a locally length-minimizing curve. In the
plane, the geodesics are straight lines. On the sphere, the
geodesics are great circles. The geodesics in a manifold under the
Finsler metric, i.e., curves in $SU(2^n)$ which are local extrema of
the Finsler length are determined by \emph{geodesic equation}, which
is a second-order differential equation.  The length of a geodesic
from I to U equals to $\int_{t=0}^{t=1} F_q(H(t))dt$. Here $H(t)$ is
a time-dependent Hamiltonian matrix satisfying Schrodinger's
equation $\frac{dV}{dt} = -i H(t)V$; $V(0)=I $ ; $V(1)=U $.

\subsection{Constant Geodesics}
The shortest path is a solution of geodesic equation. In general, it
is hard to figure out the global minima (geodesics with shortest
length) from all possible local minima. However we can study some
special geodesics corresponding to time independent Hamiltonian
systems.

\begin{deff}
\label{deff-ConstantGeodesics} Constant geodesics is a geodesic of
the form $V(t)= e^{-iHt}$ which satisfied Schrodinger's equation
$\frac{dV}{dt} = -i HV$; $V(0)=I $ ; $V(1)=U $.
\end{deff}
In this case, H corresponds to a time-independent Hamiltonian
system. The length of the constant geodesic equals to $F_q(H)$. If
we restrict the H to be a sum of commuting Pauli matrices, then it
is a \emph{Pauli geodesic}. In some condition, the global minimal
length geodesic could be a constant geodesic.

\begin{proposition}
~\cite{nielsen-2005} Let U be diagonal in the computational basis.
Suppose the minimal length geodesic $s$ between I and U is unique,
then $s$ must be a Pauli geodesic.
\end{proposition}

\section{Result and Proofs}

\begin{thm}
There exists an explicit family of unitary matrices U which has
exponentially long minimal length constant geodesics under $F_q$
metric.
\end{thm}

\begin{proof}

Set $N= 2^n$. Let $U \in SU(N)$ be a diagonal unitary matrix in the
computational basis. Suppose U has $N$ different eigenvalues
satisfying $e^{-i h_0 }$ , $e^{-i h_1 }$ ,..., $e^{-i h_{N-1} }$ and
$0 \leq h_k < 2\pi$ for $k = 0,1,...,N-1$ . Let $H = diag(h_0 , h_1
, ... , h_{N-1} )$. Let $\mathcal{J}$ be the set of diagonal
matrices $J = 2\pi diag(j_0, j_1 , ... , j_{N-1})$ where $j_k$ is an
arbitrary integer for $k = 0,1,...,N-1$. Then the whole set of
constant geodesics is $\{e^{-i(H-J)t} \hspace{1mm} | \hspace{1mm}J
\in \mathcal{J} \}$. The length is given by $F_q(H-J)$, therefore
the length of the minimal constant geodesic from I through U is
given by:
\[
    \min_{J \in \mathcal{J} } F_q(H-J)
\]

The set $\mathcal{J}$ forms an integer lattice. The problem of
finding shortest constant geodesic is equivalent to finding a
closest point in the lattice under the $F_q$ metric on group
$SU(N)$.

Intuitively, we have a $N$ dimensional Euclidean space. Dimensions
corresponding to some generalized Pauli matrix $\sigma$ which is
tensor products of some Is and Zs. A point in this space is a
diagonal Hermitian matrix. Its coordinate equals to the coefficient
of Pauli operator expansion of the matrix. This $F_q$ metric is an
anisotropic generalization of normal distance on $N$ dimensional
Euclidean space. Set $\mathcal{J}$ forms a lattice in this space.
Our goal is to find an explicit point so that the distance from this
point to the closest point in the lattice is exponential.

The lattice has some nice property which is crucial to our proofs.
If we project all lattice points to some dimension $\sigma$, the set
of points after projection is discrete and there are interval
between two consecutive projected points. Therefore we can pick some
point in the middle of this interval, which is far away from any
vertex in the lattice.

Let $col(H) =(h_0 , h_1 , ... , h_{N-1})^{T}$, A diagonal matrix can
be written as linear combination of general pauli matrices which are
tensor products of Is and Zs. We denote $col(\lambda^H)=(\lambda_0 ,
\lambda_1 , ... , \lambda_{N-1})^{T}$ as $N=2^n$ coefficients of
pauli expansion. Instead of H, we use M ($M_{ij} = (-1)^{i\cdot j})
$ to denote $\textit{Hadamard}$ matrix to avoid confusion. It is not
hard to prove following lemma:

\begin{lem}
\[ \frac{1}{N}M^{\otimes n}col(H)=col(\lambda^H)\]
\end{lem}
\begin{proof}
\[
\sum_{i}h_i |i><i| = \sum_{i} h_i (\frac{1}{N} \sum_{j} (-1)^{i\cdot
j}\sigma_j) = \frac{1}{N}\sum_{j}(\sum_{i}h_i (-1)^{i \cdot
j})\sigma_j
\]
\end{proof}

Set $H_0= \frac{\pi}{N}\sigma$, $\sigma$ is tensor product of Is and
Zs and $pw(\sigma)\geq 3$. Suppose $\sigma$ is the $i$th general
pauli matrix. We have

\begin{lem}
\[
\min_{J \in \mathcal{J} } F_q(H_0-J) \geq \frac{\pi}{N} q
\]
\end{lem}
\begin{proof}
\begin{align*}\min_{J \in \mathcal{J} } F_q(H_0-J) & \geq q \min_{J \in
\mathcal{J} } \lambda^{H_0-J}_i \\ &= q\min_{J \in \mathcal{J}}\{
\lambda^{H_0}_i - \lambda^{J}_i\} \\& = q \min_{J \in \mathcal{J}}\{
\lambda^{\frac{\pi}{N}\sigma }_i - (\frac{1}{N}M^{\otimes n}
col(J))_i \}
\\& \geq q \min_{k \in
\mathbb{Z} } \{\frac{\pi}{N} - \frac{2k\pi}{N} \}
\\&= \frac{\pi}{N} q
\end{align*}
\end{proof}

Therefore, if we set $q$ to be exponential large and we perturb $U =
e^{-i H_0 }$ a little bit to make the eigenvalues differ from each
other, we can get a $U'$, whose shortest length constant geodesics
is exponential.

\end{proof}

\section{Discussions}

In the previous section, we've showed some explicit unitary matrix
which has exponentially long minimal constant geodesic. However in
the shortest constant geodesic is not necessarily to be shortest
among all geodesics. In our example, the Hamiltonian can be
simulated by polynomial quantum circuit, therefore the the globally
minimizing geodesic is only of polynomial length. More interesting
question is following:
\begin{problem}
Can one show an explicit unitary matrix which has exponentially long
minimal geodesic?
\end{problem}
We know that if the minimal length geodesic is unique, then it must
be a Pauli geodesic. One approach to solving this problem is to find
some unitaries with exponential length constant geodesics while the
minimal length geodesic is unique. The other direction is to prove
similar results for other Pauli metrics.
\begin{proposition}
The length of constant geodesics under $F_2$ metric is no more than
$2\pi$ for any unitary matrix U.
\end{proposition}
\begin{proof}
\[
\min \{ F_2(H)  \hspace{1mm} | \hspace{1mm} e^{-iH}= U \} = \min \{
\sqrt{\frac{tr(H^2)}{N}}  \hspace{1mm} | \hspace{1mm} e^{-iH}= U \}
 \leq \sqrt{\frac{(2\pi)^2 N}{N}} = 2\pi
 \]
\end{proof}

\begin{problem}
Is there an explicit family of unitary matrices U which has
exponentially long minimal length constant geodesics under $F_1$
metric?
\end{problem}

Finally, Pauli metric seems to be related to the time complexity of
simulating Hamiltonian by quantum circuit. ~\cite{nielsen-2006-311}
If a family of unitaries U has poly length of geodesics under $F_q$
metric where q is exponential, then it can be simulated by quantum
circuits in polynomial time.
\begin{problem}
Can one show similar relation between Pauli metrics $F_1$ and time
complexity of simulating Hamiltonian system?
\end{problem}

\section{Acknowledgments}

Thanks to Pranab Sen, Sean Hallgren, Martin Roetteler and Yaoyun Shi
for stimulating discussions.


\end{document}